\documentclass[aps,superscriptaddress,prx,reprint,amsmath,amssymbols]{revtex4-2}
\usepackage{graphicx}
\usepackage{array}
\usepackage{booktabs}
\usepackage[colorlinks=true, allcolors=prxblue]{hyperref}
\usepackage{bm}
\usepackage{color}
\usepackage{subfigure}
\usepackage{soul}
\usepackage{amsmath}
\usepackage{amsfonts}
\usepackage{amssymb}
\usepackage{lipsum}
\usepackage{calrsfs}
\usepackage{tikz}
\usepackage{silence}
\usepackage[compat=1.1.0]{tikz-feynman}
\tikzfeynmanset{warn luatex=false}
\usepackage{xcolor}
\definecolor{prxblue}{RGB}{61,37,143}
\DeclareMathAlphabet{\pazocal}{OMS}{zplm}{m}{n}

\renewcommand{\arraystretch}{1.25}

\definecolor{darkBlue}{rgb}{0,0,0.6}
\definecolor{darkRed}{rgb}{0.5,0,0}
\definecolor{darkGreen}{rgb}{0,0.5,0}

\begin{document}
\title{Effective-medium theory for elastic systems with correlated disorder}
\author{Jorge M. Escobar-Agudelo}
\email{jorge.escobar@unesp.br}
\thanks{he/him/his or she/her/hers}
\affiliation{ICTP South American Institute for Fundamental Research, S\~ao Paulo, SP, Brazil}
\affiliation{Instituto de F\'isica Te\'orica, Universidade Estadual Paulista, S\~ao Paulo, SP, Brazil}
\author{Rui Aquino}
\email{ruiaquino@ictp-saifr.org}
\affiliation{ICTP South American Institute for Fundamental Research, S\~ao Paulo, SP, Brazil}
\affiliation{Instituto de F\'isica Te\'orica, Universidade Estadual Paulista, S\~ao Paulo, SP, Brazil}
\author{Danilo B. Liarte}
\email{danilo.liarte@ictp-saifr.org}
\affiliation{ICTP South American Institute for Fundamental Research, S\~ao Paulo, SP, Brazil}
\affiliation{Instituto de F\'isica Te\'orica, Universidade Estadual Paulista, S\~ao Paulo, SP, Brazil}
\affiliation{Department of Physics, Cornell University, Ithaca, NY 14853, USA}
\date{\today}
	
\begin{abstract}
Correlated structures are intimately connected to intriguing phenomena exhibited by a variety of disordered systems such as soft colloidal gels, bio-polymer networks and colloidal suspensions near a shear jamming transition.
The universal critical behavior of these systems near the onset of rigidity is often described by traditional approaches as the \emph{coherent potential approximation} --- a versatile version of effective-medium theory that nevertheless have hitherto lacked key ingredients to describe disorder spatial correlations.
Here we propose a multi-purpose generalization of the coherent potential approximation to describe the mechanical behavior of elastic networks with spatially-correlated disorder.
We apply our theory to a simple rigidity-percolation model for colloidal gels and study the effects of correlations in both the critical point and the overall scaling behavior.
We find that although the presence of spatial correlations (mimicking attractive interactions of gels) shifts the critical packing fraction to lower values, suggesting sub-isostatic behavior, the critical coordination number of the associated network remains isostatic.
More importantly, we discuss how our theory can be employed to describe a large variety of systems with spatially-correlated disorder.
\end{abstract}

\maketitle

\section{Introduction}
\label{sec:Introduction}
Disordered elastic materials on the verge of rigidity loss have routinely served as a proxy for the intricate universal critical behavior exhibited by a wide variety of systems ranging from bio-polymer networks~\cite{BroederszMac2014}, randomly packed spheres~\cite{Liu1998,Liu2010,Wyart2005} and confluent cell tissues~\cite{Bi2015} to granular media~\cite{Tkachenko1999,Edwards1999}, dislocation systems~\cite{MiguelZap2002,TsekenisDah2011} and even strange metals~\cite{liarte2022, ThorntonCho2023}.
Theoretical approaches to describe the rigidity loss of disordered solids often build upon two widespread paradigmatic transitions: Rigidity percolation~\cite{deSouza2009,Feng1984} and jamming~\cite{Liu1998,Liu2010,Wyart2005}.
Simple models for both of these transitions involve diluted versions of mass-spring networks in which there are nearly enough constraints (two-body harmonic interactions) for rigidity~\cite{Feng1984,Souslov2009,Mao2011,GoodrichNag2015,Liarte2019}.
Disorder effects are then taken into account by means of quenched independent and identically distributed (i.i.d.) random variables, e.g. bond dilution in elastic network systems~\cite{Feng1985}, exchange interactions in mean-field spin glasses~\cite{MezardVir1987}.
This approximation leads to structures that are spatially and temporally uncorrelated, and is sufficient to qualitatively (and sometimes quantitatively) describe the universal scaling behavior of many disordered elastic materials.
Yet it lacks key ingredients that are necessary to explain a number of intriguing phenomena such as the putative sub-isostatic behavior of soft colloidal gels~\cite{Zhang2019}.
Here we introduce a generalized effective-medium theory that extends the coherent potential approximation~\cite{Leath1968,Feng1985} to incorporate spatial correlations into the disordered structures of a wide class of elastic materials. We apply our theory to better understand the role played by correlations into the mechanical behavior of soft gels.

\begin{figure*}[!tb]
  \begin{center}
    \subfigure{
        \includegraphics[width=0.25\linewidth]{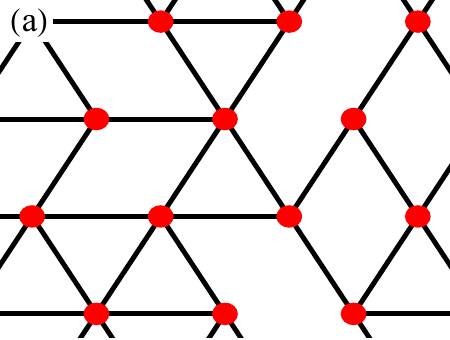}
        \label{fig:CPAa}}
    \hspace{0.03\linewidth}
    \subfigure{
        \includegraphics[width=0.25\linewidth]{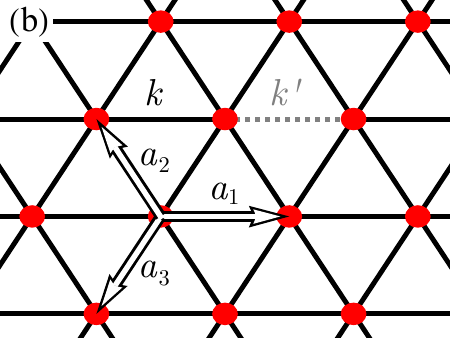}
        \label{fig:CPAb}}
    \hspace{0.03\linewidth}
    \subfigure{
        \includegraphics[width=0.25\linewidth]{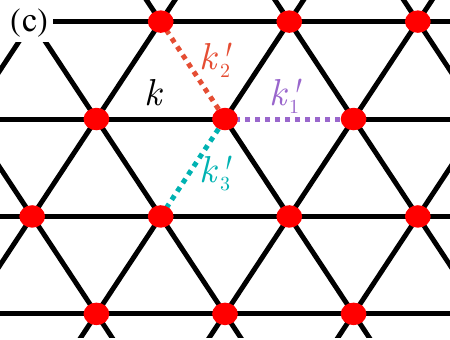}
        \label{fig:CPAc}}
    \caption{Schematic representation of the Coherent Potential Approximation.
    (a) Regular networks in which bonds (springs) are randomly removed mimics the microscopic structure of diverse disordered solids.
    (b) In traditional CPA, the randomly-diluted network depicted in (a) is mapped into a homogeneous network with effective spring constant $k$, which is determined as the solution to a self-consistent equation originating in the linear response to a perturbation on a single bond (dotted gray line).
    (c) We extend traditional CPA to include the combined effect of multiple defects, so that the elastic spring constants $k'_i$ are sampled from a distribution that captures spatial correlations.}
    \label{fig:CPA}
  \end{center}
\end{figure*}

Effective-medium theories (EMTs) have been routinely employed as a semi-analytical approach for varied disordered elastic systems.
A particularly powerful version of EMT is based on the so-called \emph{Coherent Potential Approximation} (CPA)~\cite{Leath1968}.
In the CPA, the randomly diluted network [Fig.~\ref{fig:CPAa}] is mapped into a homogeneous network [Fig.~\ref{fig:CPAb}] characterized by an effective spring constant that satisfies a self-consistent equation (see Sec.~\ref{sec:CPA}).
In the standard and most common form of the CPA, this self-consistent equation is derived from the linear visco-elastic response coming from a perturbation on a \emph{single impurity} [a bond in this case; see Fig.~\ref{fig:CPAb}].
Whereas this approximation has been successful to describe static~\cite{Feng1985,Jacobs1995,Jacobs1996,Mao2011,Liarte2020} and dynamic~\cite{DueringWya2013,YuchtBro2013,Liarte2019,liarte2022} mechanical properties of diverse systems (including systems with bond-bending forces~\cite{Das2007,Mao2013,Mao2013b,LiarteLub2016,JacksonCoh2022}), it is not capable of capturing spatial correlations of the disordered structure due to its inherent single-impurity approximation.
One of the major results in this paper is an extension of the CPA that goes beyond the single-impurity approximation [Fig~\ref{fig:CPAc}] and is capable of describing the much larger class of systems with \emph{spatially-correlated disorder}. 

\begin{figure}[!th]
\begin{center}
\includegraphics[width=\linewidth]{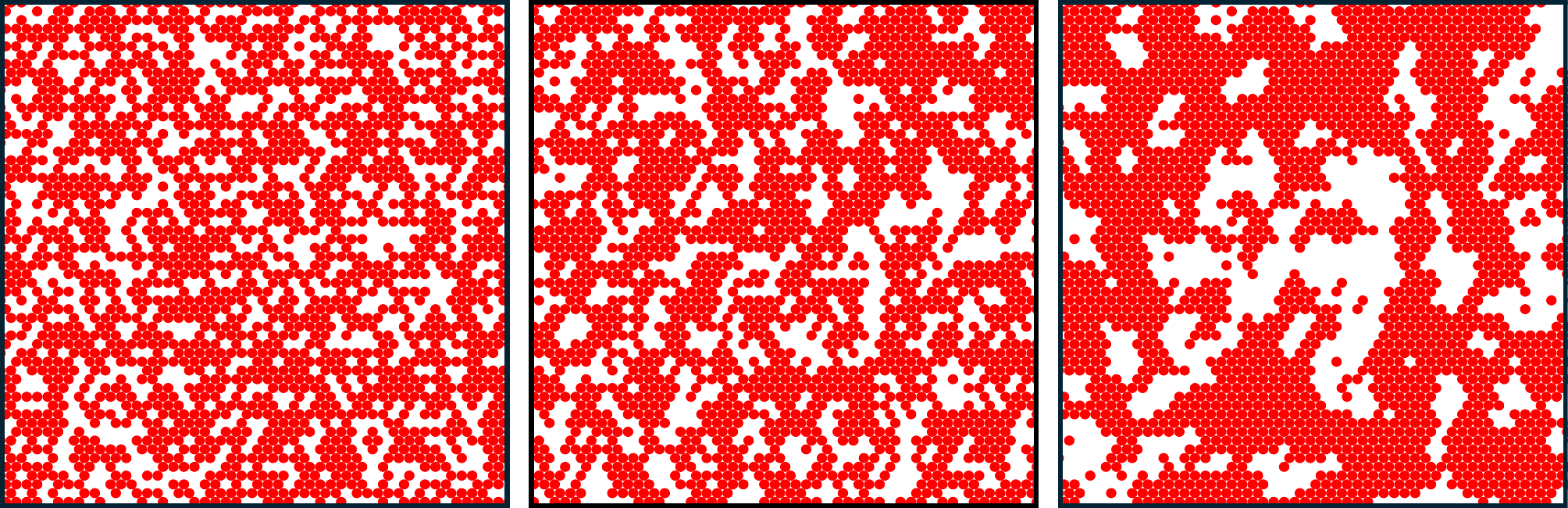}
\end{center}
\caption{Contrast between correlated and uncorrelated disordered structures.
Simulated configurations following the protocol of Ref.~\cite{Zhang2019} for a rigidity-percolation model of gels at target volume fraction $\phi_l=0.67$, and correlation strength $c=0$ (left), $0.3$ (center) and $0.6$ (right), representing a bias to add particles where there are more neighbors.
\label{fig:CorrelatedStructures}}
\end{figure}

Although much less explored, spatially-correlated disordered structures seem to be the rule, rather than the exception.
Dense colloidal suspensions under shear stress may undergo shear thickening and shear jamming~\cite{BehringerCha2018,BiBeh2011,RamaswamyCoh2023}, in which the shift from predominant hydrodynamic to frictional interactions lead to force chains and a highly-correlated disordered structure not well described by i.i.d. random variables.
Another intriguing example is that of anisotropic triangular lattices in which there is a bias for the existence of bonds along one particular direction~\cite{WangCoh2025}.
In this case, simulations show that percolation of rigid clusters happen at different critical values along different directions, a result that is not captured by a direct application of traditional EMTs.  
Finally, we analyze in detail simple network models for soft colloidal gels~\cite{Kleman2007,Richtering2014,Bouzid2018,Whitaker2019}.
Here attractive interactions lead to nontrivial structures (Fig.~\ref{fig:CorrelatedStructures}) with spatial correlations that have been linked to an apparent \emph{sub-isostatic} rigidity transition threshold~\cite{Zhang2019}. 
Our findings will show that the dominant mechanism for a lower critical packing fraction observed in correlated rigidity percolation models is mostly due to an effective increase of local coordination numbers due to correlations.
Though the critical packing fraction may be lower, the transition still happens at the \emph{isostatic} ``critical coordination point'' $z_c= 2d$~\cite{Maxwell1864,deSouza2009}, where $d$ is the spatial dimension!

This article is organized as follows. In Sec.~\ref{sec:CPA}, we review the main steps of the standard CPA formalism, as it applies to generic randomly-diluted elastic networks.
We then introduce a generalized CPA framework in Sec.~\ref{sec:GCPA}, in which spatial correlations are incorporated via the interplay of combined defects, thus going beyond the single impurity approximation.
In Sec.~\ref{sec:Gels}, we apply this theory to analyze the rigidity percolation model for gels, proposed and numerically simulated by Zhang et al.~\cite{Zhang2019}.
Finally, we make final remarks in Sec.~\ref{sec:Conclusions}.

\section{Coherent Potential Approximation}
\label{sec:CPA}
Consider a set of $N$ particles of mass $m$ that interact with their neighbors as illustrated in the networks of Fig.~\ref{fig:CPA}~\footnote{Although we focus on the particular case of a triangular lattice, our results can be easily generalized and applied to other types of network structures.}.
The position vectors of the particles are denoted by $\bf{R}_i$ and $\bf{r}_i$, for coordinates in reference (equilibrium) and target spaces, respectively.
The particles interact via harmonic elastic interactions described by the energy~\footnote{Notice that we have chosen the units of $E$ so that the lattice spacing is the unit for lengths and $g_{ij}$ is unit-less.}
\begin{equation}
E = \frac{1}{2} \sum_{\langle i, j\rangle} g_{ij} \left[\left(\bm{u}_i-\bm{u}_j\right)\cdot \hat{\bm{r}}_{ij}\right]^2,
\label{eq:Energy}
\end{equation}
where $\bf{u}_i = \bf{r}_i-\bf{R}_i$ is the displacement vector, $\hat{\bf{r}}_{ij} = (\bf{r}_j-\bf{r}_i)/\sqrt{\bf{r}_j-\bf{r}_i}$, and we assume for simplicity that the sum is restricted to nearest neighbors of a regular lattice (generalization to more complex interactions and topologies is straightforward).
Disordered structures are incorporated via the set of random variables $\{ g_{ij}\}$, which in this section are assumed to be independent and identically distributed with probability $p (g_{ij}) = p\, \delta (g_{ij}-1) + (1-p) \delta (g_{ij})$.
This is a typical framework in standard theories for rigidity transitions; as $p$ is increased beyond a critical threshold $p_c$, the system transitions from a floppy phase (with zero elastic moduli and a finite number of nontrivial zero-energy modes) to a disordered elastic phase.
Equation~\ref{eq:Energy} is more conveniently expressed in Fourier space:
\begin{equation}
E = \frac{1}{2N^2} \sum_{\bm{q},\bm{q}^\prime} \bm{u}_{\bm{q}} \cdot \bm{D}_{-\bm{q},\bm{q}^\prime} \cdot \bm{u}_{\bm{q}^\prime},
\end{equation}
where $\bm{u}_{\bm{q}}$ is the Fourier transform of $\bm{u}_i$, and the dynamical matrix can be written as
\begin{equation}
\bm{D}_{-\bm{q}, \bm{q}^\prime} = N \delta_{\bm{q}, \bm{q}^\prime} \bm{D}_{\bm{q}},
\label{eq:DM-TI}
\end{equation}
for periodic translationally-invariant systems.

The Coherent Potential Approximation (CPA)~\cite{Leath1968,Feng1985,Mao2011} is a form of effective-medium theory that maps a disordered network system [Fig.~\ref{fig:CPAa}] into a homogeneous one [Fig.~\ref{fig:CPAb}] characterized by an effective spring constant $k$ to be determined a posteriori.
In its traditional form, one introduces a \emph{single} defect on the homogeneous lattice [Fig.~\ref{fig:CPAb}], e.g. by changing one bond's elastic constant from $k$ to $k^\prime$, and calculates the lattice elastic response to this perturbation.
The constant $k$ of the effective medium is then determined by imposing that the scattering off this single impurity must be zero on average.
This can be achieved by matching the average perturbed Green's function ($\bm{G}^V$) to the unperturbed one ($\bm{G}^{(m)}$),
 \begin{equation}
     \left\langle\bm{G}^V\right\rangle=\bm{G}^{(m)},
     \label{scc}
 \end{equation}
where the average is taken over the single random variable $k^\prime$ that satisfies the probability distribution of the original network.

The zero-frequency retarded Green’s function $\bm{G}$ is related to the dynamical matrix via the equation
 \begin{equation}
     \bm{G}=-\bm{D}^{-1}
     \label{G=D^-1}.
 \end{equation}
 Here we assume a single-bond perturbation $\bm{V}$, so that the dynamical matrix changes from the translational-invariant form $\bm{D}^{(m)}$ [given by Eq.~\eqref{eq:DM-TI}] to
 \begin{equation}
    \bm{D}^V=\bm{D}^{(m)}+\bm{V},
\label{D^V=D^m+V}
\end{equation}
which in turn leads to the perturbed Green’s function
\begin{equation}
    \bm{G}^V=\left[\left(\bm{G}^{(m)}\right)^{-1}-\bm{V}\right]^{-1},
\end{equation}
which can be written as~\cite{Elliott1974} 
\begin{equation}
    \bm{G}^V=\bm{G}^{(m)}+\bm{G}^{(m)}\cdot\bm{T}\cdot\bm{G}^{(m)},
\label{Gv=G+GTG}
\end{equation}
where
\begin{equation}
\bm{T}=\bm{V}+\bm{V}\cdot\bm{G}^{(m)}\cdot\bm{V}+\bm{V}\cdot\bm{G}^{(m)}\cdot\bm{V}\cdot\bm{G}^{(m)}\cdot\bm{V}+\cdots,
\label{T-matrix expanded}
\end{equation}
is the $T$-matrix.
The CPA self-consistency condition can then be written as
\begin{equation}
\langle\bm{T}\rangle=0.
\label{eq:CPA-equation}
\end{equation}

To evaluate the average on Eq.~\eqref{eq:CPA-equation}, we need an explicit form for the perturbation $\bm{V}$ and the dynamical matrix $\bm{D}^{(m)}$, which depend on the network structure.
For the homogeneous triangular lattice of Fig.~\ref{fig:CPA}, we can write
\begin{equation}
\begin{split}
    \bm{D}_{\bm{q}}^{(m)}= k \sum_{\ell=1}^3 \bm{B}_{\ell,\bm{q}}\wedge\bm{B}_{\ell,-\bm{q}},
\end{split}
\label{D=kBB}
\end{equation}
where $k$ is the effective spring constant, the sum runs over the three bonds $\ell$ in the cell, and the bond vectors are defined as
\begin{equation}
\begin{split}
    \bm{B}_{\ell,\bm{q}}&\equiv\left( 1-e^{-i\bm{q}\cdot \bm{a}_\ell}\right)\hat{\bm{e}}_{{a}_\ell},\\
\end{split}
\label{B vector for NN case}
\end{equation} 
with $\bm{a}_\ell$ a primitive lattice vector (Fig.~\ref{fig:CPAb}) and $\hat{\bm{e}}_{{a}_\ell}=\bm{a}_\ell/|\bm{a}_\ell|$.
The symbol $\wedge$ denotes an outer product.
For a localized perturbation in one bond, the perturbation matrix becomes
\begin{equation}
\begin{split}
    \bm{V}_{\bm{q},\bm{q}'}&= (k'-k) \bm{B}_{1,\bm{q}}\wedge\bm{B}_{1,-\bm{q}'},
\end{split}
\label{V=(k'-k)BB}
\end{equation}
where $k^\prime$ is an i.i.d. random variable with distribution $p(k^\prime) = p \, \delta(k^\prime-1)+(1-p) \delta (k^\prime) $.
 Substituting Eq.~\eqref{V=(k'-k)BB} into Eq.~\eqref{T-matrix expanded}, the $T$-matrix reduces to a geometric series, yielding
\begin{equation}
    \bm{T}=\frac{\bm{V}}{1+(k'/k-1)h(k)},
\end{equation}
where
\begin{equation}
    h(k)= -v_0\, k \int_{1\text{BZ}}\frac{d^2\bm{q}}{4\pi^2}\bm{B}_{1,-\bm{q}}\cdot\bm{G}^{(m)}_{\bm{q}}\cdot\bm{B}_{1,\bm{q}}
    \label{eq:hofk}
\end{equation}
is a scalar function, with $v_0$ the unit cell volume (area in 2D) and the integral is taken over the first Brillouin zone (1BZ).
Equation~\eqref{eq:CPA-equation} then leads to the self-consistent equation for $k$,
\begin{equation}
k = \frac{p - h(k)}{1-h(k)}.
\label{eq:cpaEq}
\end{equation}
This approach has been proven highly effective in modeling randomly diluted network systems, but it inherently neglects spatial correlations, since it is based on single-impurity perturbations.
In the next section we will generalize this framework to treat correlated disorder, enabling the description of a more general class of complex systems.

\section{Correlated Coherent Potential Approximation}
\label{sec:GCPA}

To incorporate correlations we need to treat the combined effect of perturbations originating in multiple defects [Fig.~\ref{fig:CPAc}].
In other words, the perturbation matrix $\bm{V}$ must account for the simultaneous change in several bonds, each with a distinct elastic constant $k'_i$.
Correlations are incorporated by assigning an appropriate joint probability distribution for all $k^\prime_i$ in the block.
Thus, for a system with $n$ defects, we can write 
\begin{equation}
    \bm{V}_{\bm{q},\bm{q}'}=\sum_{i=1}^n(k_i'-k)\bm{B}_{i,\bm{q}}\wedge\bm{B}_{i,-\bm{q}'}.
\label{first generalization of V}
\end{equation}
Equation~\eqref{T-matrix expanded} still holds, but now the different powers of $\bm{V}$ will result in terms that mix the different bond vectors $\bm{B}_i$.
Fortunately, it is still possible to carry out calculations by considering a geometric series for matrix (rather than scalar) objects. The corresponding $T$-matrix is no longer proportional to $\bm{V}$, but takes the modified form
\begin{equation}
    \bm{T}_{\bm{q},\bm{q}'}=\sum_{i,j=1}^nk\bm{B}_{i,\bm{q}}\wedge\bm{B}_{j,-\bm{q}'}F_{ij},
\label{closed form of T matrix}
\end{equation}
where we introduce the $F$-matrix, with components
\begin{equation}
    F_{ij}\equiv\left(\frac{k_i'}{k}-1\right){\left[\bm{\delta}+\bm{\tilde{H}}\right]^{-1}}_{ij},
\label{Definition of F}
\end{equation}
where $\bm{\delta}$ is an $n\times n$ identity matrix, and the components of $\bm{\tilde{H}}$ are defined by
\begin{equation}
\tilde{H}_{ij} = \left(\frac{k_j'}{k}-1\right)H_{ij},
\end{equation}
with
\begin{equation}
\begin{split}
    H_{ij}&\equiv-k{{\int}}_{\bm{q}}d\bm{ \mathrm{q}}\ \bm{B}_{i,-\bm{q}}\cdot\bm{G}^{(m)}_{\bm{q}}\cdot\bm{B}_{j,\bm{q}}.
\end{split}
\label{definition H matrix}
\end{equation}
Notice that the matrix $H_{ij} (k)$ is a natural generalization of $h(k)$ in Eq.~\eqref{eq:hofk}.

Now the $n$ random variables $k^\prime_i$ are entirely encapsulated in the $F$-matrix, so that the average value of the $T$-matrix over all perturbations reduces to
\begin{equation}
\begin{split}
\left\langle\bm{T}_{\bm{q},\bm{q}'}\right\rangle&=\sum_{i,j=1}^nk\bm{B}_{i,\bm{q}}\wedge\bm{B}_{j,-\bm{q}'}\left\langle F_{ij}\right\rangle,\\
\end{split}
\end{equation}
and the self-consistency condition $ \left\langle \bm{T}\right\rangle=0$ yields
\begin{equation}
\begin{split}
    \left\langle \bm{F}\right\rangle=0.
\end{split}
\label{Generalized self-consistency constraint}
\end{equation}
This generalized self-consistent condition represents our main theoretical contribution --- a versatile analytical framework that goes beyond conventional mean-field approximations for random elastic systems, by capturing spatial correlations associated with more complex probability distributions.
Notice, however, that unlike the single scalar equation~\eqref{eq:cpaEq} for the determination of $k$, Eq.~\eqref{Generalized self-consistency constraint} is a matrix equation representing at most $n^2$ equation (one for each component of $\bm{F}$).
We will now see that simple symmetry arguments can be used to considerably reduce the number of independent equations (two in the case of the simple triangular lattice in Fig.~\ref{fig:CPA}).
The remaining redundancy will be addressed by invoking the single-site approximation (Appendix A).


If the chosen set of bonds have rotational symmetry [see e.g. the bonds $k^\prime_1$, $k^\prime_2$ and $k^\prime_3$ in Fig.~\ref{fig:CPAc}], it is straightforward to check that all diagonal components of the $H$-matrix are identical (i.e. $H_{11}=H_{22}=H_{33}$).
For the same reason, all off-diagonal components are identical, thus reducing the number of independent equations to just two.
Moreover, off-diagonal terms are associated with higher-order scattering processes involving clusters of defects, which can be neglected under a version of the single-site approximation (see Appendix~\ref{App:single-site approximation}).
Crucially, this approximation does not imply the elimination of spatial correlations.
We have not relaxed one of our most central conditions, namely that we need to perturb a \emph{cluster} of bonds to incorporate these correlations.
In this framework, the multiple-bond perturbation indeed plays an essential role even in the diagonal terms, which are sufficient to account for spatial correlations in the generalized probability distribution.

The diagonal components of the $H$-matrix involves the term $k\bm{B}_{i,\bm{q}}\wedge\bm{B}_{i,-\bm{q}}$, which corresponds to the $i$-th contribution to the dynamical matrix $\bm{D}^{(m)}_{\bm{q}}$.
For rotationally invariant systems, all such terms are equivalent across the Brillouin zone, allowing us to express $H_{ii}$ as
\begin{equation}
        H_{ii}=-\frac{1}{z_{NN}}\int_{\bm{q}}d\bm{ \mathrm{q}}\ \text{tr}\Big[\bm{D}^{(m)}_{\bm{q}}\cdot\bm{G}^{(m)}_{\bm{q}}\Big],
    \end{equation}
where $z_{NN}$ denotes the number of bonds per cell. 
Using Eq.~\eqref{G=D^-1}, this further simplifies to
\begin{equation}
    H_{ii}=\frac d{z_{NN}},
\label{diagonal components of H}
\end{equation}
which significantly simplifies the calculation.
This result is analogous to the one obtained for $h(k)$ in the traditional version of CPA~\cite{Feng1985}.
Lattices with a basis, or the frequency-dependent CPA will lead to more complicated $H$-matrices that can be calculated either by symbolic computation or direct numerical evaluation.

\section{Model with correlated disorder}
\label{sec:Gels}

To validate our analytical framework, we consider a system inspired by the numerical model introduced by Zhang {\it et al}. \cite{Zhang2019}.
In their model, an empty site of a 2D triangular lattice is randomly selected, and then occupied with probability
\begin{equation}
    p=(1-c)^{6-N_{nn}},
\label{Zhang probability protocol}
\end{equation}
where $N_{nn}$ ($0\leq N_{nn}\leq 6$) counts the number of occupied nearest-neighbor sites, and $c \in [0,1]$ tunes the correlation strength ($c=0$ corresponds to the uncorrelated case).
The process is iterated until a target volume fraction $\phi_l$ is reached.
Neighbor pairs of occupied sites form a bond (a spring) that is described by a typical harmonic elastic interaction. 
The simulations show that rigidity transitions shift from the expected isostatic threshold [$\phi_l (c=0)$] to lower values as $c$ increases --- an indication of apparent sub-isostatic behavior.

\begin{figure}[!t]
\begin{center}
\includegraphics[width=\linewidth]{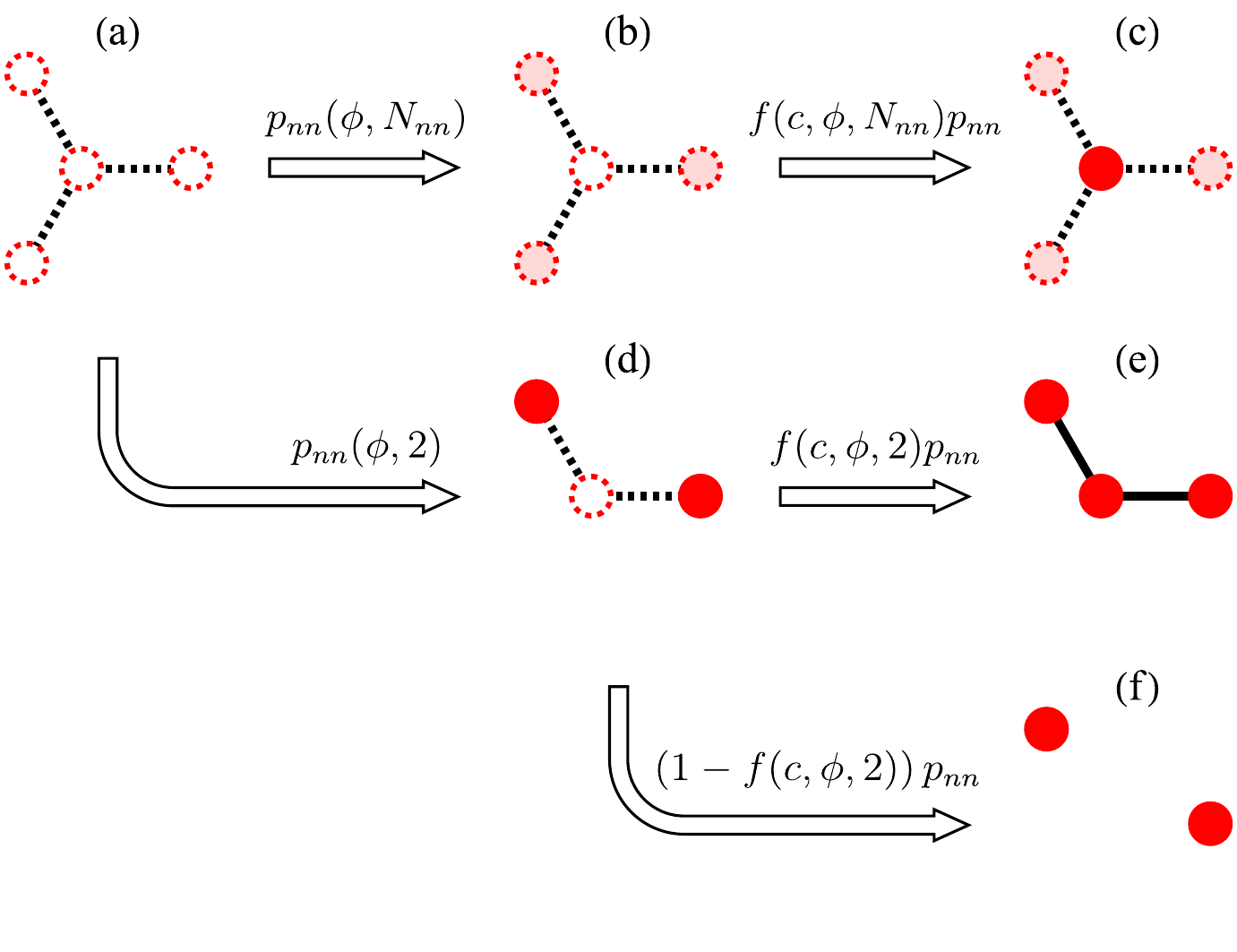}
\end{center}
\caption{Rigidity-percolation model for gels in the CCPA framework.
(a) We start with a triangular lattice unit cell with no occupied bonds and no occupied sites.
(b) A set of $N_{nn}$ particles is then added at the outer sites with uncorrelated probability $p_{nn}$.
(c) Finally, a particle is added at the center with \emph{correlated} probability $f(c, \phi, N_{nn})$.
Panels (d) - (f) show an example in which two particles are added at the outer sites (d), followed by occupation or not of the central site [(e) and (f), respectively].
\label{fig:phi model}}
\end{figure}

Here we consider an adaptation of the model of Ref.~\cite{Zhang2019} amenable to analytical calculations based on our correlated coherent potential approximation (CCPA).
To apply our method, we need to assign a joint probability for configurations associated with each set of bonds in the unit cell.
For this, we start with the triangular lattice unit cell shown in Fig.~\ref{fig:phi model}a, in which neither bonds nor sites are occupied.
We then add particles to each of the outer sites with occupation probability $\phi$, so that a configuration with $N_{nn}$ particles has probability (Fig.~\ref{fig:phi model}b)
\begin{equation}
    p_{nn}=p_{nn}(\phi,N_{nn})=\phi^{N_{nn}}(1-\phi)^{3-N_{nn}}.
\end{equation}
Notice that at this stage correlations have not yet been incorporated.
This can be achieved by adding a particle at the center with probability
\begin{equation}
    f(c) =  f(c, \phi, N_{nn}) = A\, (1-c)^{3-N_{nn}},
\end{equation}
which has the same form as Eq.~\eqref{Zhang probability protocol}.
The constant $A$ is chosen to enforce the average occupation of the lattice to be $\phi$, which leads to the form
\begin{equation}
A = A (\phi, c) = \frac{\phi}{\left[1+c(\phi - 1)\right]^3}.
\end{equation}
It is straightforward to check that $c=0$ corresponds to the uncorrelated case in which a particle is added at the center with independent probability $\phi$.
The joint probability for a configuration with $N_{nn}$ outer sites occupied is then given by (Fig.~\ref{fig:phi model}c)  
\begin{equation}
    p=\begin{cases} 
    p_{nn}f(c), & \text{if the center is occupied,}\\
    p_{nn}(1-f(c)), & \text{otherwise.}
\end{cases}
\end{equation}
Finally, each pair $(i,j)$ of occupied neighbors contributes a harmonic elastic energy of the standard form $(1/2) [(\bm{u}_i-\bm{u}_j)\cdot \hat{\bm{r}}_{ij}]^2$.
In this way, a configuration of particles, with its respective probability, is directly mapped into the configuration of bonds that we need for the implementation of the CCPA [Eq.~\eqref{Generalized self-consistency constraint}].
Figures~\ref{fig:phi model}d-f show an example of our protocol.
A configuration with two occupied outer sites happen with probability $p_{nn} (\phi, 2)$ (d).
The central site is then occupied (e) or unoccupied (f) with probabilities $f(c, \phi, 2) p_{nn} (\phi, 2)$ and $[1-f(c, \phi, 2)] p_{nn} (\phi, 2)$, respectively.

\begin{figure}[!ht]
\begin{center}
\includegraphics[width=\linewidth]{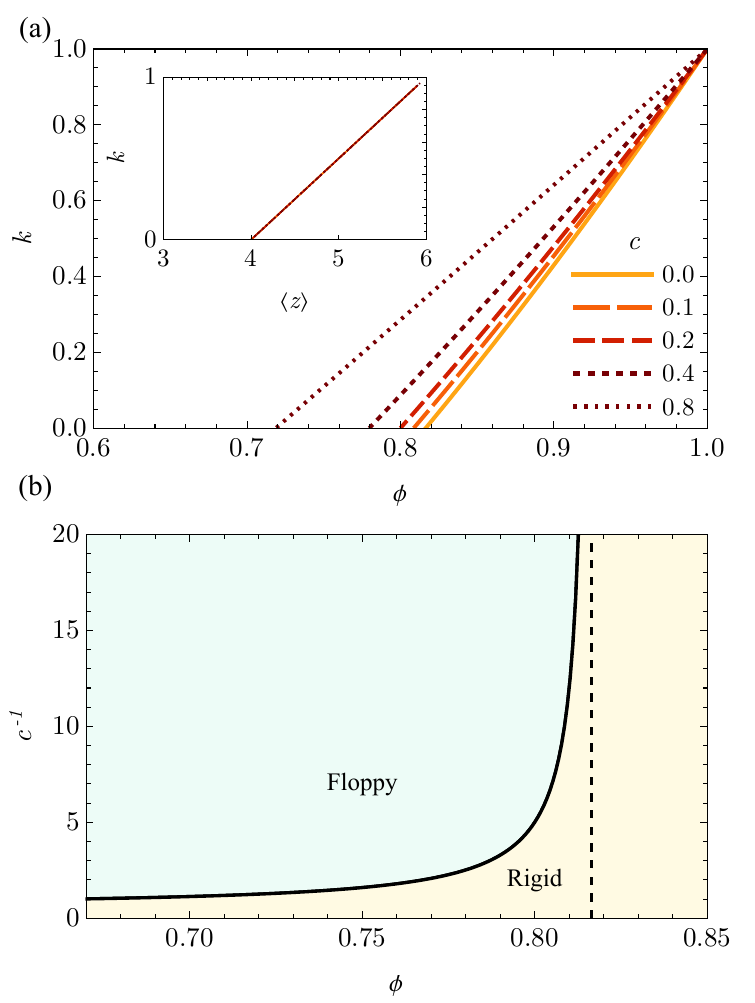}
\end{center}
\caption{(a) Effective elastic constant $k$ as a function of packing fraction $\phi$ for several values of the correlation strength $c$.
The uncorrelated case ($c=0$, solid line) shows standard rigidity-percolation behavior, whereas the transition of correlated systems ($c>0$, dashed lines) are shifted to lower values of $\phi$.
The inset shows that all curves collapse when $k$ is plotted as a function of average coordination number $\langle z \rangle$.
(b) Phase diagram showing in terms of packing fraction $\phi$ and the inverse of the correlation strength $c^{-1}$.
The continuous black line represent the phase boundary between the floppy and the rigid states and is obtained analytically using the generalized coherent potential approximation.
\label{fig:kvsphi_phase-diagram}}
\end{figure}

Now we solve the generalized CCPA equation [Eq.~\eqref{Generalized self-consistency constraint}] to study the dependence of the effective elastic constant $k$ on both packing fraction $\phi$ and correlation parameter $c$.
Figure~\ref{fig:kvsphi_phase-diagram}a shows a plot of $k$ as a function of $\phi$ for several values of the correlation parameter $c$.
For $c=0$ (full yellow line), i.e. the uncorrelated case, we recover standard rigidity percolation behavior, with $k=0$ below a critical volume fraction $\phi_c(c=0) =\sqrt{2/3}$, and a continuous increase for $\phi>\phi_c$ until $k=1$ at $\phi=1$.
In this limiting case our homogeneous lattice exactly maps the original system.
For the correlated case $c>0$, we observe the expected monotonic decrease of the critical packing fraction, which suggests an apparent sub-isostatic behavior.
Nevertheless, as the inset shows, all curves collapse into a single curve when $k$ is plotted as a function the average coordination number $\langle z \rangle$.
This result shows that the transition is (strictly speaking) \emph{isostatic} --- The rigidity transition occurs at the expected Maxwell threshold $z_c=2 d$.
Above this point, the effective spring constant linearly increases with $\langle z\rangle$, reaching $k=1$ at $\langle z\rangle=6$ of the homogeneous triangular lattice.
We will further explain and discuss this intriguing behavior in the next paragraph.
Figure~\ref{fig:kvsphi_phase-diagram}b shows a phase diagram in terms of $c^{-1}$ and $\phi$.
Our analytical results (black curve) are in excellent agreement with the numerical simulations of Ref.~\cite{Zhang2019} (the discrepancy in the value of $\phi_c(c=0)$ is due to a slightly different definition for $\phi$; we use the ratio of occupied sites whereas they use the ratio of occupied volume).

To gain a better understanding of the role of correlations on isostaticity, we will further explore our CCPA equations using an analytical calculation (which is still possible for this simple model in the triangular lattice).
The average coordination number, $\langle z\rangle$, can be written as
\begin{equation}
\begin{split}
  \langle z \rangle &= \langle z\rangle(c,\phi) = 2 \big[3 f(c,\phi,3) p_{nn}(\phi,3)
   \\& \quad + 
   2 (3f(c,\phi,2) p_{nn}(\phi,2)) + 
   1 (3f(c,\phi,1) p_{nn}(\phi,1))\big] \\& \quad 
   =  \frac{6\phi^2}{1+c(\phi-1)}.
\end{split}
\label{calculation of z}
\end{equation}
In turn, our generalized self-consistency equation yields
\begin{equation}
    k(c,\phi)=\frac{3\phi^2}{1+c(\phi-1)}-2.
\label{form of k}
\end{equation}
Equations~\eqref{calculation of z} and~\eqref{form of k} lead to the simple linear relation
\begin{equation}
    k(c,\phi)=\frac{1}{2}\langle z\rangle(c,\phi)-2,
\end{equation}
which shows that the rigidity transition happens at the Maxwell threshold in 2$d$: $z_c=4$.
Hence, the shift to lower critical packing fractions at finite $c$ cannot be simply attributed to the presence of correlations.
Instead, at fixed packing fraction, increasing the correlation strength $c$ leads to an increase in the average coordination number.
This result follows directly from Eq.~\eqref{calculation of z}, but can also be intuited from the snapshots shown in Fig.~\ref{fig:CorrelatedStructures} --- by giving a bias to fill empty sites surrounded by more neighbors, one inadvertently effectively increases $\langle z \rangle$.
Ultimately, the excellent agreement between our analytical calculations and the numerical simulations of Ref.~\cite{Zhang2019} suggests that this effective increase is the main mechanism responsible for the decrease of $\phi_c$ as a function of the correlation strength $c$.

\begin{figure}[!ht]
\begin{center}
\includegraphics[width=1\linewidth]{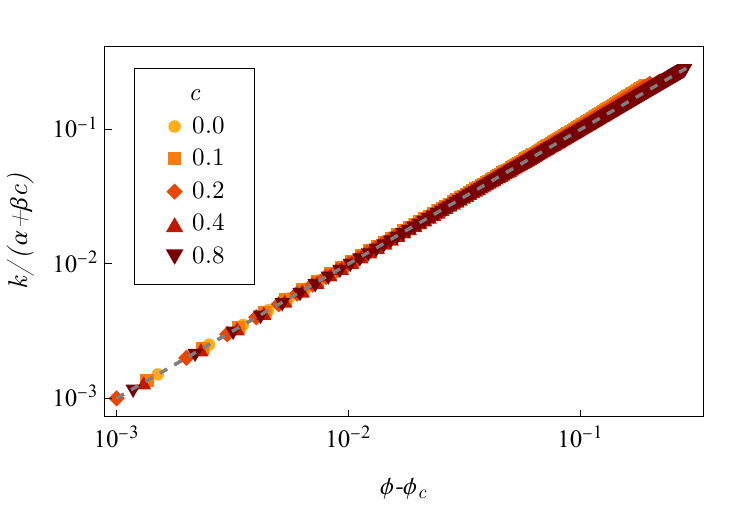}
\end{center}
\caption{Scaling collapse plot in terms of $k / (\alpha + \beta c)$ and $\phi - \phi_c$, for several values of the correlation parameter $c$ and packing fraction $\phi$. All curves collapse onto a straight line (dashed line) in the vicinity of the critical packing fraction.}
\label{fig:scale invariance}
\end{figure}

We finish this section with a simple analysis of the scaling behavior of $k=k(\phi-\phi_c,c)$.
At low values of $c$ and near the rigidity threshold $\phi_c$, Eq.~\eqref{form of k} results in
\begin{equation}
    k\approx(\alpha+\beta c)(\phi-\phi_c)
\end{equation}
where $\alpha=2\sqrt6$ and $\beta=\sqrt6-4$.
Fig.~\ref{fig:scale invariance} shows a scaling collapse plot in terms of $k/(\alpha+\beta c)$ and $\phi-\phi_c$, for several values of $\phi$ and $c$.
Notice that all curves for different values of $c$ collapse into the single linear curve. 

\section{Final Considerations}
\label{sec:Conclusions}

Here we have developed a versatile generalization of the coherent potential approximation to describe elastic systems with spatially-correlated disordered structures. 
Our approach is analytically tractable and applicable to the much larger class of \emph{correlated} disordered elastic materials that include soft gels, colloidal suspensions, bio-polymer networks and granular systems, among others.
In particular, we have reproduced simulation results for a simple rigidity percolation model of gels~\cite{Zhang2019}, and have shown that the observed decrease of the critical packing fraction as correlation strength increases is not solely due to the presence of correlations --- it mostly follows from an effective increase of average coordination number.

Whereas the mathematical framework developed in this paper is generally well posed, its successful application depends on a few important factors. 
Notice that the generalized self-consistent condition [Eq.~\eqref{Generalized self-consistency constraint}] is a matrix equation corresponding to at most $n^2$ equations, where $n$ is the number of perturbed bonds in the theory.
We have shown that for the simple triangular lattice considered here, for a cluster with $n=3$ bonds, there are just two (out of the nine) independent equations.
We would then need two variables in order to have a closed system of equations.
One possibility considered by us was to include additional interactions in the effective medium (e.g. interactions between next-nearest neighbors) described by an effective coupling to be self-consistently determined along with $k$ in our CCPA equations.
This approach did not produce sensible physical results, which suggested that the choice of effective interactions was not appropriate for the particular correlated system considered here.
We then invoked the single-site approximation (see Appendix~\ref{App:single-site approximation}), which is already inherited in the traditional CPA, to justify ignoring higher-order scattering processes represented by the off-diagonal terms in Eq.~\eqref{Generalized self-consistency constraint}.
An important extension of our theory would suitably incorporate these processes while maintaining a closed system of self-consistent equations.

It would be interesting to apply our formalism to other systems with complex disorder configurations.
A natural extension of our results for the rigidity-percolation model of gels would be a systematic investigation of the effects of correlations over longer length scales.
To this aim, one would need to consider larger unit cells in the triangular lattice, with suitable symmetry, for which one could define a spatially-decaying correlation length.
Potentially these results could shed light into connections between decaying length scales and softness, as well as the intriguing hierarchical structures of particulate colloidal gels~\cite{BantawaGad2023}.
Other possible applications include dense colloidal suspensions under shear~\cite{BehringerCha2018}, anisotropic rigidity percolation~\cite{WangCoh2025} and topological mechanics~\cite{MaoLub2018}.
From a more fundamental theoretical perspective, it would be interesting to investigate the interplay between spatially-correlated disorder and the renormalization group flows of rigidity transitions using the framework developed in Ref.~\cite{ThorntonLia2025}.

\begin{acknowledgments}
We thank Xiaoming Mao, Emanuela Del Gado and Anton Souslov for useful comments and suggestions. 
We also thank financial support by FAPESP through Grants No. 2021/14285-3 and No. 2022/09615-7 (D.B.L.),  No. 2023/00067-0 (J.M.E-A), and No. 2023/05765-7 (R.A.)
\end{acknowledgments}

\appendix

\section{Single-site approximation}
\label{App:single-site approximation}

In this appendix, we provide a justification for the single-site approximation by examining the structure of the perturbation matrix $\bm{V}$ and its implications for phonon propagation.
We neglect defect-defect interactions in the $F$-matrix and retain only its diagonal components, so that we are able to reduce the analysis to the case of single-defect scattering processes.
This approximation preserves the essential physics of the problem through the self-consistency constraint, while fully incorporating correlations via the generalized probability distribution. 
Crucially, it excludes higher-order scattering processes (e.g., pairs or clusters of defects).

The perturbation matrix (Eq.~\eqref{first generalization of V}) is a sum of independent single-defect terms:
\begin{equation}
\begin{split}
    \bm{V}_{\bm{q},\bm{q'}}=\sum_{i=1}^n\bm{V}_i(\bm{q},\bm{q}'),
\end{split}
\end{equation}
where $\bm{V}_i(\bm{q},\bm{q}')$ corresponds to a defect at bond $i$.
While $\bm{V}$ itself contains no crossed terms, the $T$-matrix expansion introduces defect interactions through higher-order terms:
\begin{equation}
\begin{split}
    \bm{T}_{\bm{q},\bm{q}'}&=\sum_{i=1}^n\bm{V}_i(\bm{q},\bm{q}')\\
    &+\sum_{i,j=1}^n\int_{\bm{q}_1}d \bm{\mathrm{q}}_1 \bm{V}_i(\bm{q},\bm{q}_1)\cdot\bm{G}_{\bm{q}_1}\cdot\bm{V}_j(\bm{q}_1,\bm{q}')+\cdots.
\end{split}
\label{T for the diagrams}
\end{equation}
To adopt a single-site approximation, we retain only terms with $i=j$ , thus neglecting interactions between distinct defects.
This is enforced statistically by requiring that different $V_i$ have zero correlation, i.e., 
\begin{equation}
\begin{split}
    \left\langle\bm{V}_i\bm{V}_j\right\rangle=c_{i}\bm{\delta}_{ij},
\end{split}
\end{equation}
where $c_i$ represents the correlation for the self-product $\bm{V}_i\bm{V}_i$ and $\bm{\delta}_{ij}$ ensures that only diagonal contributions survive.
The average is taken over disorder configurations.
The approximation is further clarified if we use the diagrammatic notation shown in Table~\ref{tab:diagrammatic_notation}.

\begin{table}[h]
    \centering
    \renewcommand{\arraystretch}{1.5}
    \begin{tabular}{c c}
        \toprule
        \textbf{Matrix} & \textbf{Diagrammatic Notation} \\
        \midrule
        \( \bm{V}_i(\bm{q},\bm{q}') \) & \raisebox{-0.45cm}{
\begin{tikzpicture}
  \begin{feynman}
  
    \vertex[dot, minimum size=2pt] (11) at (0,0.5) {};

    \draw[scalar, draw=black, line width=0.8pt] (0,0) -- (11);

    \node[below=-1pt] at (0,0) {\footnotesize $i$};
    
  \end{feynman}
\end{tikzpicture}
}\\
        \( \bm{G}_{\bm{q}_1} \) & \raisebox{-0.1cm}{
\begin{tikzpicture}
  \begin{feynman}

    \draw[scalar, draw=black, line width=0.8pt] (0,0) -- (1,0);

    \node[above=-1pt] at (0.5,0) {\footnotesize $1$};
    
  \end{feynman}
\end{tikzpicture}
} \\
        \bottomrule
    \end{tabular}
    \caption{Diagrammatic notation for the relevant matrices in the analysis.  The defect-induced scattering matrix $\bm{V}_i(\bm{q},\bm{q}')$ is represented by a vertical line, while the Green’s function $\bm{G}_{\bm{q}_1}$, describing phonon propagation at wavevector $\bm{q}_1$, is represented by a horizontal line.}
    \label{tab:diagrammatic_notation}
\end{table}

The $T$-matrix expansion, represented diagrammatically as
\begin{equation}
\bm{T}_{\bm{q},\bm{q}'} = \sum_{i=1}^n\!
\raisebox{-0.65cm}{
\begin{tikzpicture}
  \begin{feynman}
  
    \vertex[dot, minimum size=2pt] (11) at (0,0.8) {};

    \draw[scalar, draw=black, line width=0.8pt] (0,0) -- (11);

    \node[below=-1pt] at (0,0) {\footnotesize $i$};
    
  \end{feynman}
\end{tikzpicture}
}\!+\sum_{i,j=1}^n\! \raisebox{-0.71cm}{
\begin{tikzpicture}
  \begin{feynman}
  
    \vertex[dot, minimum size=2pt] (01) at (0,0.8) {};
    \vertex[dot, minimum size=2pt] (11) at (0.4,0.8) {};
    
    \draw[scalar, draw=black, line width=0.8pt] (0,0) -- (0.4,0);
    \draw[scalar, draw=black, line width=0.8pt] (0,0) -- (01);
    \draw[scalar, draw=black, line width=0.8pt] (0.4,0) -- (11);

    \node[below=-1pt] at (0,0) {\footnotesize $i$};
    \node[below=-1pt] at (0.4,0) {\footnotesize $j$};
    \node[above=-1pt] at (0.2,0) {\footnotesize $1$};
    
  \end{feynman}
\end{tikzpicture}
}\!+\sum_{i,j,l=1}^n\! \raisebox{-0.71cm}{
\begin{tikzpicture}
  \begin{feynman}
  
    \vertex[dot, minimum size=2pt] (01) at (0,0.8) {};
    \vertex[dot, minimum size=2pt] (11) at (0.4,0.8) {};
    \vertex[dot, minimum size=2pt] (22) at (0.8,0.8) {};
    
    \draw[scalar, draw=black, line width=0.8pt] (0,0) -- (0.4,0);
    \draw[scalar, draw=black, line width=0.8pt] (0,0) -- (01);
    \draw[scalar, draw=black, line width=0.8pt] (0.4,0) -- (11);
    \draw[scalar, draw=black, line width=0.8pt] (0.4,0) -- (0.8,0);
    \draw[scalar, draw=black, line width=0.8pt] (0.8,0) -- (22);

    \node[below=-1pt] at (0,0) {\footnotesize $i$};
    \node[below=-1pt] at (0.4,0) {\footnotesize $l$};
    \node[below=-1pt] at (0.8,0) {\footnotesize $j$};
    \node[above=-1pt] at (0.2,0) {\footnotesize $1$};
    \node[above=-1pt] at (0.6,0) {\footnotesize $2$};
    
  \end{feynman}
\end{tikzpicture}
}+\cdots,
\end{equation}
reduces, after taking the average, to irreducible diagrams,
\begin{equation}
\langle\bm{T}_{\bm{q},\bm{q}'}\rangle = \sum_{i=1}^n\!
\raisebox{-0.65cm}{
\begin{tikzpicture}
  \begin{feynman}
  
    \vertex[dot, minimum size=2pt] (11) at (0,0.8) {};

    \draw[scalar, draw=black, line width=0.8pt] (0,0) -- (11);

    \node[below=-1pt] at (0,0) {\footnotesize $i$};
    
  \end{feynman}
\end{tikzpicture}
}\!+\sum_{i=1}^n\! \raisebox{-0.71cm}{
\begin{tikzpicture}
  \begin{feynman}
  
    \vertex[dot, minimum size=2pt] (11) at (0.2,0.8) {};
    
    \draw[scalar, draw=black, line width=0.8pt] (0,0) -- (0.4,0);
    \draw[scalar, draw=black, line width=0.8pt] (0,0) -- (11);
    \draw[scalar, draw=black, line width=0.8pt] (0.4,0) -- (11);

    \node[below=-1pt] at (0,0) {\footnotesize $i$};
    \node[below=-1pt] at (0.4,0) {\footnotesize $i$};
    \node[above=-1pt] at (0.2,0) {\footnotesize $1$};
    
  \end{feynman}
\end{tikzpicture}
}\!+\sum_{i=1}^n\! \raisebox{-0.71cm}{
\begin{tikzpicture}
  \begin{feynman}
  
    \vertex[dot, minimum size=2pt] (11) at (0.4,0.8) {};
    
    \draw[scalar, draw=black, line width=0.8pt] (0,0) -- (0.4,0);
    \draw[scalar, draw=black, line width=0.8pt] (0,0) -- (11);
    \draw[scalar, draw=black, line width=0.8pt] (0.4,0) -- (11);
    \draw[scalar, draw=black, line width=0.8pt] (0.4,0) -- (0.8,0);
    \draw[scalar, draw=black, line width=0.8pt] (0.8,0) -- (11);

    \node[below=-1pt] at (0,0) {\footnotesize $i$};
    \node[below=-1pt] at (0.4,0) {\footnotesize $i$};
    \node[below=-1pt] at (0.8,0) {\footnotesize $i$};
    \node[above=-1pt] at (0.25,0) {\footnotesize $1$};
    \node[above=-1pt] at (0.55,0) {\footnotesize $2$};
    
  \end{feynman}
\end{tikzpicture}
}+\cdots,
\end{equation}
corresponding to the truncated series
\begin{equation}
\begin{split}
    \bm{T}_{\bm{q},\bm{q}'}&=\sum_{i=1}^n\bm{V}_i(\bm{q},\bm{q}')\\
    &+\sum_{i=1}^n\int_{\bm{q}_1}d \bm{\mathrm{q}}_1 \bm{V}_i(\bm{q},\bm{q}_1)\cdot\bm{G}_{\bm{q}_1}\cdot\bm{V}_i(\bm{q}_1,\bm{q}')+\cdots.
\end{split}
\end{equation}
This yields the diagonal form of the $H$-matrix,
\begin{equation}
\bm{H}=\frac{d}{z_{NN}}\mathbb{I},
\label{simplified definition of H after single-site approximation}
\end{equation}
which matches Eq.~\eqref{diagonal components of H}.

Substituting this result into the definition of the $F$-matrix in Eq.~\eqref{Definition of F} yields the explicit form
\begin{equation}
    F_{ij}=\frac{\delta_{ij}}{1-\frac{2}{3}\left(k'_i-k\right)}.
\end{equation}
Notice the diagonal structure of $\bm{F}$ reflects the absence of scattering between distinct defects.
Nonetheless, the dependence of $\bm{F}$ on $k'_i$ allows us to represent all possible configurations of a disordered, correlated system.
After taking an average over disorder---through the probability distribution governing the $k'_i$---we ensure the effective medium retains statistical signatures of correlations despite the single-site approximation.
This allows our analytical framework to capture correlations through the mean-field response determined by the $F$-matrix and enforced by the generalized self-consistency condition.

Notice that the inclusion of defect-defect scattering processes would lead to an overdetermined system, which could be handled with the inclusion of additional effective parameters beyond $k$.
Whereas the single-site approximation provides a justified simplification for systems where correlations between defects are negligible, the absence of higher-order defect interactions might be relevant for e.g. strongly correlated systems.
Therefore a careful assessment of these limitations has to be made before the application of our methods to these systems.

\end{document}